\documentclass{llncs}
\usepackage[T2A]{fontenc}
\usepackage[utf8]{inputenc}
\usepackage{verbatim}
\usepackage{textcomp}
\usepackage{color}
\usepackage{amssymb}
\usepackage{latexsym}
\usepackage{amsmath}

\newcommand{\EQ}{\mathrm{EQ}}
\newcommand{\AC}{\mathrm{AC}}

\begin{document}
\title{Amortized communication complexity of an equality predicate.(Beta version)}
\author{Vladimir Nikishkin}
\institute{Moscow Institute of Physics and Technology}
\date{2012}
\maketitle

\begin{abstract}
We study the communication complexity of  the direct sum of independent copies of the equality predicate. We prove that the probabilistic communication complexity of this problem is equal to $O(N)$; computational complexity of the proposed protocol is polynomial in size of inputs. Our protocol improves the result achieved in \cite{feder}. Our construction is based on two techniques: Nisan's pseudorandom generator \cite{nisan:generator}  and Smith's string synchronization algorithm  \cite{smith}.
\end{abstract}

\section{Introduction}

In this paper we study amortized communication complexity of the equality predicate. We deal with the classic model of communication complexity with two participants (Alice and Bob), who want to compute some function of the data distributed between the participants. Alice and Bob can talk to each other via a communication channel. We measure the number of bits that must be transmitted between Alice and Bob to achieve the goal.

More specifically, let $f : \{0,1\}^n\times\{0,1\}^n\to \{0,1\}$ be a function of two arguments. We assume that Alice is given the value of $x$, Bob is given the value of $y$, and Alice and Bob communicate with each other to compute the value $f(x,y)$. We denote by $C(f)$ deterministic communication complexity of function $f$, i.e., the minimal number of bits that should send by Alice and Bob to each other to get $f(x,y)$.  

Further, let us denote by $f^N$ the direct sum of $N$ independent copies of the initial function $f$. More formally, the two arguments of $f$ are an $N$-tuples of values
$(x_1,\ldots, x_N)$ and $N$-tuple of values $(y_1,\ldots,y_N)$, and 
$$
f^N(x_1,\ldots, x_N, y_1,\ldots,y_N) = (f(x_1,y_1),\ldots,f(x_N,y_N)).
$$
We assume that Alice is given all values of $x_i$, and Bob is given all values of $y_i$. Now ALice and Bob need to compute $f^N$, i.e., to get all the values $f(x_i,y_i)$ for $i=1,\ldots,N$.
Amortized communication complexity of  $f$ is defined as 
$$
\AC(f) = \limsup\limits_{N\to \infty} \frac{f^N}{N}.
$$
It was proven in \cite{feder} that 
			$AC(f) = \Omega(\sqrt{C(f)} - \log{n})$.
			
In a similar way, amortized communication complexity can be defined for probabilistic version of communication complexity. In general, the properties of  amortized randomized communication complexity remains not well understood, though several  nontrivial particular examples are known. For instance, in \cite{feder} it was proven that amortized randomized complexity of the \emph{equality predicate} is only $O(1)$, while it is known that for one individual predicate
$\EQ:\{0,1\}^n\times \{0,1\}^n\to\{0,1\}$
$$
\EQ(x,y) = \left\{
\begin{array}{rl}
1,&\mbox{if }x=y,\\
0,&\mbox{if }x\not=y
\end{array}
\right.
$$
randomized communication complexity is equal to $\Theta(\log n)$.
		
In our paper we construct a new randomized communication protocol for the direct sum of the equality predicate.
Our protocol is computationally effective, i.e.,  it only requires polynomial computations for Alice and Bob.
Our construction improves the result from \cite{feder} in two respects. First of all, we get a slightly better bound for the probability of  error. And, second, our protocol has a ``modular'' structure; it consist of several  independent gadgets, which make the construction more flexible. We hope that the same technique can be applied to other problems of communication complexity.

		Our principal result can be formulated as follows:

\medskip

\noindent
\textbf{Main theorem.}
Probabilistic communication complexity (for the private coin model) of a direct sum of $N$ equality predicates is equal to $O(N)$, with probability of the error $P_{err} \leq O(2^{-c\frac{N}{\log^2 N}})$. Moreover, we explicitly construct a communication protocol that achieves this communication complexity  and requires only polynomial time computations on Alice's and Bob's sides.

\medskip

In our construction, we use several classic tools (N.~Nisan's pseudorandom generator, BCH codes,  deterministic synchronization protocol by A.~Orlitsky) and one relatively new construction (A.Smith's probabilistic synchronization protocol).

		
	\subsection{Model}
	
We use three standard models of communication complexity: deterministic communication protocols, randomized communication protocols with public random bits, and randomized communication protocols with private random bits,
see  Nisan's and Kushlevitz's textbook~\cite{book}. We done communication complexities for these three models by
$C_{det}$, $C_{pub}$, and $C_{priv}$ respectively.

Our principal construction is in a sense, an explicit and effective implementation  of the following theorem for some particular communication protocol.
\begin{theorem}[Comm. Compl. \cite{book}]\label{privatization}
		Let 
		$f: \{0, 1\}^n \times \{0, 1\}^n \rightarrow \{0, 1\}$ 
be a function of two arguments. For every $\delta > 0$ and every $\varepsilon > 0$, it holds
$$C_{priv}^{\varepsilon+\delta}(f) < C_{pub}^\varepsilon(f) + O(\log n + \log \delta^{-1}).$$ 
\end{theorem}

	\subsection{Our main problem}

		\textbf{Equality predicate ($EQ_n$).} First of all, we remind the following classic problem of communication complexity.
		Alice and Bob each hold an n-bit string, $x$ and $y$ respectively. They want to know whether $x=y$. Formally, they 
		want to compute the value of the predicate
		$$
		EQ_n(x,y) =\left\{
		\begin{array}{rl}
		  1, & \mbox{if }x=y,\\
		   0, & \mbox{if }x\not=y.
		 \end{array}  
		\right.
	 $$
	 
\textbf{Direct sum of equality functions ($EQ_n^N$).}
Both Alice and Bob hold some array of $N$  of $n$-bits blocks. Formally, inputs of Alice and Bob are bit strings of length $n\cdot N$. But it is more instructive to represent the input of Alice  as $x=x^1\ldots x^N$ and the input of Bob  as $y=y^1\ldots y^N$, where $x_i, y_i\in\{0,1\}^n$ for each $i=1,\ldots,N$. Then we define $EQ_n^N$ as
		 $EQ_n^N(x,y) = z \in \{0,1\}^N$, where each $i$-th bit of $z$ is equal to $1$ iff $x^i=y^i$. 
Intuitively	this means that Alice and Bob wants to compute the value of the predicate $EQ_n$ for $N$ independent pairs of inputs $(x^i,y^i)$.
		
Communication complexity of the function $EQ_n^N$ is the main subject of our paper. More precisely, we want to estimate the probabilistic communication complexity of this function in the model with private sources of randomness.
		
\section{The known results}
	\subsection{Complexity of $EQ_n$ for different types of communication protocols.}
The predicate $EQ_n$ is pretty well studied, and its communication complexity are well understood. Let us remind three different communication protocols for this predicate.
		
\paragraph{Deterministic model.} It is known that
			$C_{det}(EQ_n) = n+1.$
The bound is achieved by a trivial protocol: Alice transmits her string $x$ to Bob, Bob compares the two strings $x$ and $y$ and sends back one-bit response, $1$ if the strings are equal and  $0$ otherwise. From the standard technique of fooling sets it follows that this bound is tight, i.e., there is no protocols with communication complexity less than $n+1$.
		
\paragraph{Private coin model}\label{eqsol2} For the randomized communication complexity with private sources of randomness 
			$C_{priv}^{\varepsilon}(EQ_n) = O( \log \frac{n}{\varepsilon}).$
This bound is achieved by several classic communication protocols.	In what follows we describe one of them.
Alice and Bob view their inputs $x$ and $y$ as $n$-digits binary representations of integers (between $0$ and $2^n-1$ ). Alice chooses a prime number $p$ at random among the first $(n/\varepsilon)$ primes. She sends to Bob both $p$ and $x\mod p$. Bob verifies whether $x \mod p = y \mod p$. If $x$ and $y$ are equal to each other modulo $p$, then Bob returns outputs $1$, otherwise he returns $0$. 

If $x=y$, then this protocol always return the correct result. If $x\not=y$, then the difference $(x-y)$ has at most $n$ prime factors; hence, the protocol returns the wrong answer with probability at most $\varepsilon$.

\paragraph{Public coin model}\label{eqsol3} For the randomized communication complexity with public sources of randomness 
			$$C_{pub}^{\varepsilon}(EQ_n) = O(\log \frac{1}{\varepsilon}).$$
This bound for the communication complexity is achieved by the following protocol.	Alice and Bob jointly choose a random $n$-bit string $n$. Then Alice computes the inner product $b = \langle x, r\rangle$ and transmits the result (a single bit) to Bob. Bob checks whether $b = \langle y, r\rangle$ and outputs "equal" if so and "not equal" otherwise. Obviously, if $x = y$, then the output is always "equal." On the other hand, if $x \neq y$, then by the properties of the inner product, $Pr[ \langle x, r\rangle \neq \langle y, r\rangle] = \frac{1}{2}$. Thus, Bob outputs "not equal" with probability $\frac{1}{2}$. To decrease the probability to get the wrong answer, Alice and Bob should repeat these procedure several times with several independently chosen random strings $r$. If Alice and Bob repeat (in parallel or sequentially) $l$ times the discribed procedure,  then the probability that  $\langle x, r\rangle \neq \langle y, r\rangle$ for all $r_1,\ldots, r_l$ is equal to $2^{-l}$. So, for $l=\lceil \log 1/\varepsilon\rceil $ we reduce the probability to get an error to $\varepsilon$, while communication complexity is $O(\log 1/\varepsilon)$.
		
	\subsection{Trivial generalizations for $EQ_n^N$}
The protocols from the previous section can be easily adapted to get some protocols for the direct sum of $N$ copies of $EQ_n$, i.e., for the function $EQ_n^N$.
		
\paragraph{Adaptation of the protocol from paragraph  \ref{eqsol2}}
			We run the protocol independently for each pair of blocks$(x_i, y_i)$. The probability to get a wrong answer \emph{for at least one pair of blocks}  must be bounded by $\varepsilon$. To this end we need to reduce the probability of an error for each of $N$ pairs of blocks to be less than $\epsilon'=\varepsilon/N$. This results in communication complexity $ O( N(\log (n/\varepsilon'))) = O( N(\log n+ \log N + \log 1/\varepsilon)))$. Thus, from the trivial adaptation of the protocol from paragraph  \ref{eqsol2} we get
					$$C_{priv}^{\varepsilon}(\EQ^N)= O( N(\log n+ \log N + \log 1/\varepsilon)))$$
		
\paragraph{Adaptation of the protocol from paragraph \ref{eqsol3}}
We run the protocol from section \ref{eqsol3} for each pair of blocks $(x_i, y_i)$ independently. To guarantee than the total probability of the error is bounded by $\varepsilon$, we need to reduce the probabilities of errors for each pair of blocks to $\epsilon'=\epsilon/N$. Then we get
					$$C_{pub}^{\epsilon}(\EQ^N)= O( N(\log N + \log 1/\epsilon)))$$
		
\paragraph{From public to private randomness}
The last protocol above can be transformed into a protocol with private sources of randomness.
Indeed, from theorem~\ref{privatization} we get immediately
	$$C_{priv} = O(N\cdot \log N + \log{\frac{1}{\epsilon}} + \log(n\cdot N) + \log{\frac{1}{\delta}}) =$$
	$$=O(N\cdot \log N + \log n +\log N + \log{\frac{1}{\epsilon}}) = O(N\cdot \log N )$$
Note that this communication protocol requires exponential computational complexity (at least for the standard proof of theorem~\ref{privatization}).

Can we reduce the obtained (rather trivial) bound $O(N\cdot \log N )$, hopefully to $O(N)$? Can we achieve this bound with a communication protocol that requires only poly-time computations?
The answers both these questions are positive. Construction of such a communication protocol is the main result of this paper. Loosely speaking, we plan to do it in two steps. At the first step, we construct more effective communication protocol for communication model with public randomness (this part of our construction is based on ideas of A.~Smith). At the second step, we reduce this protocol with a public source randomness to a protocol with private randomness. In some sense, this idea is similar to the usual proof of theorem~\ref{privatization}: we substitute the sequence of random bits (shared by Alice and Bob) by a sequence of \emph{pseudorandom} bits, which can be obtained as an output of a pseudo-random bits generator. A random seed of this generator is rather short. So, one of participants can choose it at random and then send to another participants. E.g., Alice choses a random seeds and send it to Bob; then ALice and Bob apply the pseudo-random bits generator to this same seed, and then both participants share the same long string of pseudo-random bits. The sharp difference between our construction and the standard general proof of theorem~\ref{privatization} is that we use an explicit and effectively computable generator (the generator of N.~Nisan).

Before we explain details of our construction, we remind the technical tools used in our proof.
		
\section{The tools used in our construction}
	\subsection{Pseudorandom number generator}
	In our construction we need a pseudo-random generator that fools tests with a bounded memory. Technically, we assume that a generator is a mapping $G : \{0,1\}^m \rightarrow \{0,1\}^n$, and a test is a randomized Turing  machine with working space of some size $S$.

\textbf{Definition.} \label{nisgen}
		A function $G : \{0,1\}^m \rightarrow \{0,1\}^n$ is called a pseudorandom generator is $\varepsilon$-robust for tests with space $S$, if for every \emph{statistical test} $A$ with $S$ bits of working space  
		$$|Pr_{y\in_r\{0,1\}^n}[A  \mbox{ accepts } y]- Pr_{y\in_r\{0,1\}^m}[A \mbox{ accepts }G(x)]| < \varepsilon.$$ 
By \emph{statistical test} we mean the following construction.

\textbf{Definition}\label{testmachine} A statistical test with space $S(n)$ is a deterministic Turing machine $M$ with three tapes: a working tape of size space $S(n)$, an auxiliary read-only tape with some binary strings $a= (a_1,... a_n,...)$ (an advice string), and a one-way input tape with an $n$-bits input $x$ (the reading head  on the input tape can move from the left to the right but cannot move back to the left). We always assume that the length of $a$ should not be greater than $exp(S(n))$.
This machine returns input $1$ or input $0$. We denote this result $M_a(x)$. Informally the output means that test accepts/rejects $x$ given an advice string $a$.

Nisan suggested in \cite{nisan:generator} an explicit construction of a pseudo-random generator that fools tests with small enough memory. This result of Nisan can be formulated as follows. 
\begin{theorem}
There exists a constant $c > 0$ such that for any $R$ and $S$ there exists a pseudorandom generator  $G : cS\log R \to R$  (computable in time $poly(R)$)  that is $2^{-S}$-robust for statistical tests with $S$ bits of working space.
\end{theorem}
		
In section~\ref{crack} we construct some statistical test, which verifies that a (pseudo)random string $x$ is suitable for our communication protocol. Then, we use the standard argument: our protocol with high probability returns the correct answer when it runs with truly random public bits; futher, the generator of Nisan fools our test; hence, given In section~\ref{crack} we construct some statistical test that tests that a random string is suitable for our communication protocol. Then, we use the standard argument: the protocol with high probability returns the correct answer if runs on truly random public bits; futher, the generator of Nisan fools our test; hence, given pseudo-random bits instead of truly random ones, the  communication protocol must also returns the correct answer with hight probability.

			\subsection{BCH codes}
	\label{bch}
We also use in our construction the classic BCH-codes, see \cite{bch}. We do not employ any specific properties of the construction of the BCH codes. We use only the fact that  
		$\forall m >3, t<2^{m-1}$ exists an explicit construction of a linear code with parameters $[n,k,d]$ such that codeword length $n = 2^m-1 $, the number of checksum bits is $n-k \leq mt$, and the minimal distance between of the code is $d\geq 2t+1$. We also use the fact that the BCH codes can be decoded efficiently (by Berlekamp-Messy algorithm), \cite{bch-dec}. 
		
		The BCH construction is not explicitly used in the article. Still, Orlitzky's construction \ref{orlalg} we do utilize, needs an error-correcting code, which is not explicitly given in Orlitzky's article. The BCH code is fits in his construction and we are going to use it when refering to Orlitzky's construction.

	\subsection{Strings synchronization protocols} 
In our communication protocol we will need to solve the following auxiliary problem.
Let Alice and Bob each hold an $n$-bits string,  $A$ and $B$ respectively. We assume that $A$ and $B$ differ from eahc other in at most $e$ positions. Alice and Bob want to exchange their inputs, i.e., Alice should get string $B$, and Bob should get string $A$.  We will call this problem by \emph{string synchronization problem}	(Alice and bob want to synchronize their inputs).

Orlitsky suggested in \cite{orlitsky} \label{orlalg}
a deterministic communication protocol for the problem of synchronization of a pair $n$-bits strings at the Hamming distance at most $e$. Communication complexity of this protocol is $O(e\log{n})$. All computations of Alice and Bob in this protocol run in polynomial time. 

More formally, the theorem (formulation taken from \cite{chuklin}) looks the following:
\begin{theorem}
Given an error-corrcting code with parameters $(\alpha. R(\alpha))$ which satisfies the following conditions:

\begin{enumerate}
\item{It is linear.}
\item{An effective decoding algorithm exists.}
\end{enumerate}

A one-round communication protocol with communication complexity $C = (1 - R(\alpha))\cdot n$ can be constructed. Computational complexity of such protocol is polynomial.
\end{theorem}

If using the BCH code(noted in section \ref{bch}), the communication complexity of this protocol is:
$O(e\log{n})$\\
The protocol of Orlitsky makes sense if the distance $e$ between strings is very small. In case $e=\Omega(n)$, communication complexity of the protocol of Orlitsky is worse than the trivial bound $2n$.

\label{smialg}
Adam Smith suggested in  \cite{smith} a randomized communication protocol for the problem of strings synchronization, with asymptotically  optimal bound for communication complexity for the case $e = const \cdot n$.  
More precisely, Smith proved that  for  every $\delta = \delta(n) =\Omega( \frac{\log \log n}{\sqrt{\log n}} )$ there exists an explicit family of communication protocols (with private sources of randomness) that solve the problem for synchronization of $n$ bit strings at that differ in at most $e$ positions, with communication complexity $n(H(\frac{e}{n}) + \delta)$ and error $\varepsilon = 2^{-\Omega(\frac{\delta^3n}{ \log n})}$, where 
 $H(p)=p\log_2\frac1p+(1-p)\log\frac1{1-p}.$
 Algorithms of Alice and Bob in this protocol run in polynomial time.

\section{The main result}
\label{myresults}
	\subsection{Formulation}

By $EQ_n$ we denote an equality predicate problem, by $EQ_n^N$ we denote a direct sum of such problems.

\begin{theorem}
Probabilistic communication complexity (for the private coin model) of a  $EQ^N_n$ is equal to $O(N)$, with an error probability $P_{err} \leq O(2^{-c\frac{N}{\log^2 N}})$ if n < N. Moreover, there exists a protocol with the required communication complexity and only polynomial time computations for Alice and Bob.

\end{theorem}

	\subsection{Overview of the protocol}
		Our protocol runs as follows. 
		First of all, Alice generates a  string of truly random bits of length $O(N)$ and send this string to Bob. They both generate pseudo-random bits from this seed.   In what follows, Alice and Bob use this long string of pseudo-random bits.
		
Then, Alice and Bob iteratively calculate "checksums" (inner products with pseudo-random string) for their $n$-bits blocks and synchronize strings of resulting checksums using the probablilistic or the deterministic protocol from section~\ref{orlalg}. As soon as some pair of non-equal blocks $X^i$, $Y^i$ is revealed (if some checksums for these blocks are different), Alice and Bob withdraw these blocks from the list of their bit strings and never test them again. Thus, on each next iteration the fraction of  non-equal pairs of blocks (that are not discovered yet) becomes less and less.

On  each next iteration, we make the length of  checksums  longer and longer, so for each pair of non-equal blocks the probability to be discovered becomes closer and closer to $1$. Hence, the fraction of (non-discovered) pairs of non-equal blocks gradually reduces, and only pairs of equal blocks remain untouched at their places. This means that  on each next iteration the Hamming distance between arrays of checksums (obtained by Alice and Bob respectively) becomes less and less. 

On each iteration Alice and Bob need to exchange the checksums computed for their blocks of bits (inner products with the same pseudo-random bits). For several first iterations (technically, for $\log \log N$ iterations)  we use the randomized  synchronization protocol by Smith. Then we switch to the deterministic protocol by Orlitsky.
		
In what follows we explain this protocol in more detail.
		
		\subsubsection{Generation stage}\label{genstage}
		
Alice generates $$r=\log{(n\cdot N)^4} \cdot \log( 2^{\frac{N}{\log{(n\cdot N)}}}) = O(N)$$
random bits and sends them to Bob. Then Alice and Bob apply Nisan's pseudorandom generator from section~\ref{nisgen} and get $R=n^2N^2$ pseudo-random bits. 
The length of the seed $r$ is chosen so that the generator is $\varepsilon$-robust against tests with working space of size $\frac{N}{\log{(n\cdot N)}}$.
	
		\subsubsection{Probabilistic synchronization stage, steps $i=1,\ldots,\log \log N$)}\label{probsync}
		
Synchronization protocols we use expect to know the distance between strings in advance. As we may not know an initial distance between X and Y, we will add N dummy, equal blocks to X and Y. This will guarantee the share of non equal pairs to be less then 0.5. This is a coarse trick, but it will not affect the asymptotic complexity of our protocol.
		
We repeat $\log \log N$ times the following procedure. We let $\lambda_i = \frac{2^i}{\log N}$.
Alice and Bob calculate checksums of length $\lambda_i$ for each of their blocks (that are not yet proven to be different). The checksum for each block consists of inner products modulo  $2$ between this block and a new portion of  pseudorandom bits  generated on the previous stage. Thus, the resulting checksums (for Alice and Bob) consists of $\lambda_i N$ bits.
	
Then Alice and Bob exchange their checksums using the randomized protocol of strings synchronization; when we apply  this protocol, we assume  that  Alice's and Bob's checksums differ from each other in a fraction at most  $2^{-i}$. 

When the checksums are exchanged, Alice and Bob remove from their lists the blocks $X^i$, $Y^i$ whose checksums are not identical.

Note that for a pair of equal blocks $X^i$, $Y^i$, the checksums are always equal. If blocks are not equal to each other, the chance to get all equal checksums is about $2^{-\lambda_i}$ (this probability is not \emph{exactly}  $2^{-\lambda_i}$ since Alice and Bob use not random but pseudo-random bits to compute the inner products).

Typically, on each step the number of non-discovered pairs of non-equal blocks $X^i$, $Y^i$	 becomes more than twice less. We say that the $i$-th step of the described procedure \emph{fails}, if at this stage Alice and Bob discover less than $50\%$ of the pairs of non-equal blocks $X^i$, $Y^i$ (less than a half of all pairs of non-equal blocks that was not discovered earlier). If at least one step fails, we cannot guarantee correctness of the result of the protocol. If no steps fail, then on each $i$-th step the arrays of checksums of Alice and Bob differ from each other in a fraction at most $1/2^i$ of all computed inner products.

Communication complexity of this stage is the sum of communication complexities of copies of Smith's protocol run for each step $i=1,\ldots,\log \log N$:
						$$\sum_{i=1}^{\log{\log{N}}} H(1/2^i)\lambda_i N=O(N).$$ 
The last equation follows from the choice of $\lambda_i$ and the asymptotic $H(\alpha)=\alpha \log(1-\alpha)+O(1)$ as $\alpha$ tends to $0$. 
		
		\subsubsection{Deterministic synchronization stage, steps $i=\log \log N+1,\ldots,\log{N}$}\label{detsync}
			
At	this stage we continue essentially the same procedure as at the prevoius stage. On each step Alice and Bob get $\lambda_i = \frac{2^i}{\log^2 N}$ bits of random checksums for each pair of blocks (that are not proven yet to be not-equal) by computing the inner products with new portions of pseudo-random bits; then Alice and Bob exchange the computed checksums.
The difference is only how the participants exchange their checksums. Now they use the deterministic protocol by Orlitsky instead of the instead of probabilistic protocol by Smith, see section~\ref{smialg}.
			
Communication complexity of the deterministic protocol is about $\log N$ times grater than complexity of the protocol by Smith. But nevertheless  we can use it since the value of $\lambda_i$ is reasonably small. The communication complexity of this stage is
			$$\sum_{i=\log{\log{N}}}^{\log{N}} [\frac{N \lambda_i \log N}{ 2^i}]=\sum_{i=\log{\log{N}}}^{\log{N}} {\frac{ \log N \cdot 2^i}{\log^2 N \cdot 2^i}\cdot N} = O(N).$$

		\subsubsection{Summary}
		
When the described stages are completed, we believe that Alice and Bob has found all pairs of non-equal blocks.  In all remaining pairs $X^i$, $Y^i$ (in all pairs of blocks whose checksums at all steps of the protocol were equal to each other) are considered equal. 
			
	\subsection{Probability of an error}\label{errcalc}
	We need to estimate the probability of an error in our protocol.
	For simplicity, let us assume at first, that instead of $R$ pseudo-random bits Alice and Bob share $R$ independent and uniformly distributed random bits (so, we temporarily switch to the model with a public source of randomness).
Then stages~\ref{detsync} and~	\ref{detsync} make sense, and we can estimate the probability of an error in the protocol.

		The protocol may return a wrong answer because of the following reasons:
		\begin{enumerate}
		\item The probabilistic synchronization protocol of Smith's fails at some stage. 
		\item Some of steps $i=1,\ldots,\log{N}$ fails since to many random checksums are equal for non-equal pairs of blocks $X^i$, $Y^i$. 
		\end{enumerate}
Let us bound probabilities of each of these bad events.

		\subsubsection{Error in the probabilistic synchronization.}
		
	We  sum up the probabilities of errors in Smith's synchronization at each step of our protocol:
			$$P(Err) = \sum_{i=1}^{\log \log N} O(2^{-(\frac{N}{ \log N})}) \leq O(2^{-\frac{cN}{ \log N }})$$
for some constant $c>0$.

		\subsubsection{The failure of because of checksums}
Some step $i=1,\ldots,\log{N}$ fails if for more than a half of (not discovered yet) pairs of non-equal blocks $X^i$, $Y^i$ all random checksums turn out to be equal.  We estimate the probability of this event  with Chernoff's inequality. We may assume that after the first $(i-1)$ steps there remain $N/2^i$ pairs of pairs of non-equal blocks.

We use our checksum scheme like some kind of a filter. That is - we "test" each pair of blocks on equality using calculated checksum of length $\lambda$. If the pair consists of non-equal blocks, then the test either successfully discoveres it (with probability $1- 2^{-\lambda}$) or not. The failure happens if less then a half of tests succeed. We estimate the probability of failure using Chernoff's inequality.

Formulating differently:
On each step i, the $\frac{N}{2^i}$(amount of undiscovered pairs) tests are performed, each failing with probability $2^{-\lambda_i}=$. The whole step fails if more than a half of the tests fail.

For one step, this probability can be estimated:

			$$
			 P({\mbox{Filtering out less then 1/2 pairs}}) < 2^{\frac{N}{2^i}D(q,p)}
			$$
			Where $$D(q,p) = q\ln{(\frac{q}{p})} + (1-q)\ln{(\frac{1-q}{1-p})}, \mbox{ and } q = \frac{1}{2}, p = \frac{1}{2^{\lambda_i}}$$
			Substituting $\lambda \approx \frac{2^i}{\log^2 N}$, this error probability is less then
			$$P(Err_i) \leq O(2^{-\frac{N\lambda_i}{ 2^{i}}})  = O(2^{-\frac{N}{\log^2 N}})$$

			Summing error probabilities for all steps of stages 2 and 3:
			$$\sum_{i=1}^{\log{N}} O(2^{-\frac{N}{\log^2 N}}) \approx \log{N}\cdot O(2^{-\frac{N}{\log^2 N}}) \approx O(2^{-\frac{N}{\log^2 N}+\log{\log N}})  \leq O(2^{-c\frac{N}{\log^2 N}})$$

		\subsubsection{Pseudorandom generator}\label{crack}

In this section we construct a statistical test (see the definition in section \ref{testmachine}) that  simulates one step of our protocol. In a sense, this test verifies that (pseudo)random bits are ``suitable'' for our communication protocol: they do not cause the failure of the protocol at the $i$-th iteration. 

The "advice strings" of this statistical test contains a sequence of pairs of blocks $X^i$, $Y^i$ from the inputs Alice and Bob, that were not shown to be not equal  before iteration $i$. The test should work correctly for all advice strings that correspond to the possible  internal states of Alice and Bob at the beginning of iteration $i$.

The input $x$ is a string of (pseudo)random bits that should be accepted or rejected. The test must reject $x$ (for some advice string $a$), if our communication protocol ``fails'' at the $i$-th iteration with this random bits $x$ while Alice and Bob are given the blocks $X^i$, $Y^i$  corresponding to the  advice $a$.

The algorithm of the test is straightforward: it computes the checksums for $X^i$ and $Y^i$ as it is done by our communication protocol at the $i$-th iteration, with random bits $x$ shared by Alice and Bob, and compares the corresponding checksums for Alice's and Bob's blocks.  Note that the test  does not simulate the synchronization procedure (the sub-protocols following the construction of Orlitsky and Smith).

The working space of our machine is $O(\frac{N}{\log^2 N})$. This is enough to simulate the computation of the checksums performed by our communication protocol.
The test accepts $x$, if in the simulation at least $50\%$ of non-equal pairs of blocks are  is successfully revealed, and rejects $x$ otherwise. In other words,  a teststring $x$ is rejected if it causes a failure at the $i$-th iteration of the protocol.

Theorem~\ref{nisgen} guarantees that Nisan's pseudo-random generator fools this test. Hence, for our protocol the probability of failure with pseudo-random bits is not much greater than the probability of failure for truly random bits.
More precisely, difference between the probabilities of  failure for random and pseudo-random bits is at most
$$2^{-S} = O(2^{-c \frac{N}{\log^2 N} }).$$
Technically, we should sum up this difference for all steps $i=1,\ldots,\log N$, but this does not change the asymptotics:
$P(Err) \leq \sum_{i=1}^{\log N} O(2^{-\frac{N}{\log^2 N}}) = O(2^{-c_3\frac{N}{\log^2 N}})$
	
		\subsubsection{Summary}
		
Probability of the error of our protocol consists of three parts: (1) probability of an error in Smith's protocol, (2) probability of failure with truly random checksums on some step $i=1,\ldots,\log N$, and (3) the additional probability of failure caused by the difference between truly random and pseudorandom checksums:
					$$P(Err) = O(2^{-c_1(\frac{N}{ \log N})}) + O(2^{-c_2\frac{N}{\log^2 N}}) + O(2^{-c_3\frac{N}{\log{N}}})=O(2^{-C\frac{N}{\log^2 N}}).$$
This concludes the proof of correctness of our communication protocol.
	Compared to the previous results(presented in \cite{feder}): $P(Err) = O(2^{-\sqrt{N}})$, our protocol provides better probability of error.

\end{document}